\documentclass[12pt]{article}
\usepackage{graphicx}
\usepackage{color}
\newcommand{\beq}{\begin{equation}}
\newcommand{\eeq}{\end{equation}}
\newcommand{\bea}{\begin{eqnarray}}
\newcommand{\eea}{\end{eqnarray}}

\newcommand{\gsim}{\lower.7ex\hbox{$
\;\stackrel{\textstyle>}{\sim}\;$}}
\newcommand{\lsim}{\lower.7ex\hbox{$
\;\stackrel{\textstyle<}{\sim}\;$}}

\newcommand{\eod}{\end{document}}

\def\cp{{\bf CP}}
\def\cpt{{\bf CPT}}

\definecolor{verm}{rgb}{0.8,0.1,0.0}

\begin{document}
\thispagestyle{empty}
\vspace*{-22mm}

\begin{flushright}

UND-HEP-14-BIG\hspace*{.08em}01\\

\end{flushright}

\vspace*{1.3mm}

\begin{center}
{\Large {\bf CP Asymmetries in Three-Body Final States in \\
Charged $D$ Decays \& CPT Invariance}}

\vspace*{10mm}

{\em From Roman history about data: Caelius (correspondent of Cicero) had taken a pragmatic judgment of  who was likely to win the conflict and said: 
“Pompey had the better cause, but Caesar the better army, and so I became a Caesarean.”}
\footnote{`Cause' = symmetry, yet `army' = data.} 

\vspace*{10mm}

{\bf I. Bediaga$^a$, I.I.~Bigi$^b$, J. Miranda$^a$, A. Reis$^a$} \\
\vspace{7mm}
$^a$ {\sl  Centro 
Brasileiro de Pesquisas F\'\i sicas, Rua Xavier Sigaud 150, 22290-180, 
Rio de Janeiro, RJ, Brazil}\\
$^b$ {\sl Department of Physics, University of Notre Dame du Lac}\\
{\sl Notre Dame, IN 46556, USA}

\vspace*{-.8mm}

{\sl email addresses: bediaga@cbpf.br, ibigi@nd.edu, 
jussara@cbpf.br, alberto@cbpf.br}

\vspace*{10mm}

{\bf Abstract}\vspace*{-1.5mm}\\
\end{center}
The study of regional \cp \ asymmetry in Dalitz plots of charm (\& beauty) decays 
gives us more information about the underlying dynamics than the ratio between 
total rates. In this paper we explore the consequences of the constraint from
\cpt \ symmetry with emphasis on three-body $D$ decays. We show simulations of 
$D^{\pm} \to \pi^{\pm}K^+K^-$ and discuss correlations with measured 
$D^{\pm} \to \pi^{\pm}\pi^+ \pi^-$. There are important comments about  
analysies of recent LHCb data in 
\cp\ asymmetries for $B^{\pm}$ decays to three-body final states.

\vspace{3mm}

\hrule

\tableofcontents
\vspace{5mm}

\hrule\vspace{5mm}

\section{Probing CP asymmetries with CPT invariance}

The study of \cp \ violation (CPV) is a portal to New Dynamics (ND). Although
no obvious signal of ND has been shown in hadronic data,
there are still good reasons  for its existence: neutrinos 
oscillations, the existence of Dark Matter \& Dark Energy, and `our own existence'
are the most obvious ones. 

CPV is a well established phenomenon in decays of $K$ and $B$ mesons, but
no \cp \ asymmetry has been found in $D$ decays -- yet.
The Standard Model (SM) predicts very small CPV in singly Cabibbo suppressed (SCS) 
$D$ decays, and close to zero  CPV in doubly Cabibbo suppressed (DCS) ones. 
The observation of  \cp \ asymmetries at $\mathcal{O}(10^{-2})$ level in charm decays 
would be a clear manifestation of ND. The experimental sensitivity, however, is rapidly 
reaching $\mathcal{O}(10^{-3})$ with no signals of CPV. Although the observation of CPV in 
charm would be a great achievement in itself, one would still have the difficult problem of 
disentangling ND effects from SM ones.

The vast majority of experimental searches and theoretical works refer to {\em two}-body 
final states (FS) of the type $D \to P_1P_2$ ($P$ denotes light pseudoscalar mesons). From
the theory side there are large uncertainties related to the hadronic degrees of freedom
that could easily hide the impact of ND. From the experimental side, the usual \cp\ asymmetries 
in {\em two}-body decays give `only' a single number and this is not enough information
to understand the nature of an eventual CPV signal. One needs to go beyond {\em two}-body decays 
and look at new observables. Three- and four-body decays are the natural way. First, local effects may 
be larger than phase space integrated ones. The asymmetry is modulated by the strong phase variation 
characteristic of resonant decays  \cite{BIANCO}. The   \cp \  asymmetry may change sign across the phase 
space, and the comparison between integrated rates would dilute an eventual signal. Furthermore,
the pattern of the   \cp \  asymmetry across the phase space could give insights about the underlying
operators. CPV searches with charged $D$ mesons with three-body FS is therefore a very important 
programme.

In this paper we investigate the possible patterns of CPV in three-body, SCS non-leptonic decays 
of $D$ mesons. ND could produce sizable asymmetries in DCS decays -- but DCS rates are small. 
We focus on direct CPV with $\Delta C =1$ forces and explore the correlations introduced by strong 
final state interactions (FSI) \& \cpt \ conservation, which is assumed to be exact.

Some comments are in order: 

\begin{itemize}

\item 
Theoretical predictions are more complicated in $D \to PV$ with narrow vector meson resonances $V$,
since one deals with three-body FS and interference with other intermediate states must be taken into
account. It comes much more complicated in $D \to PS$ with broad  scalar resonances $S$ -- in particular 
with $\sigma$ \& $\kappa$. From the experimental side, a full Dalitz plot with very large data sets is
quite challenging: for example, effects such as FSI among the three hadrons must 
be included in the decay model. The modeling of the S-waves is another instance of limitations of the
existing tools.

\item One alternative are model independent searches, comparing directly the $D^+$ and the $D^-$ Dalitz 
plots, as in \cite{MIRANDA1,MIRANDA2}, which is a convenient first step.
  
\item Additional constraints should be used, such as \cpt \  invariance. It comes into `play'  
by imposing equalities of total decay rates of particle and antiparticle. Its invariance imply also 
the equalities of partial rates of classes of FS where their members can re-scatter to each other. Given that
three-body decays are mostly a sequence of two-body transitions, pocesses such 
as $2 \pi , K \bar K \leftrightarrow 2\pi , K \bar K$ and  $\pi K \leftrightarrow \pi K$ become crucial.  In other
words, \cpt \  invariance would relate asymmetries in $D \to \pi\pi\pi$ with 
$D\to \pi K \bar K$.

\end{itemize} 

 \cpt \  invariance is a clear instance where the low energy hadron dynamics play an important role: it is
an unavoidable ingredient to the decay amplitude models. However, going from quarks to hadrons and understanding 
the dynamics of three-body FSI are real challenges in a quantative way. Dispersion relations and chiral perturbation 
theory are some of the theoretical tools needed for a more realistic description of three-body FS. This is indeed a field 
plenty of opportunities.

\subsection{\cp\ asymmetries \& \cpt\ constraints} 
\label{CPTCON}
 
Let us consider a decay into a FS $f$ that can proceed through two different amplitudes:
\bea
T(D^+ \to f) &=& e^{i\phi_1^{\rm weak}}e^{i\delta_1^{\rm FSI}}{\cal A}_1+ 
e^{i\phi_2^{\rm weak}}e^{i\delta_2^{\rm FSI}}|{\cal A}_2|
\\
T(D^- \to \bar f) &=& e^{-i\phi_1^{\rm weak}}e^{i\delta_1^{\rm FSI}}{\cal A}_1 + 
e^{-i\phi_2^{\rm weak}}e^{i\delta_2^{\rm FSI}}{\cal A}_2
\eea
In charged $D$ mesons only {\em direct} \cp \ violation is possible, which means 
$\Gamma (D^+ \to f) \neq \Gamma (D^- \to \bar f)$. Computing the \cp\ asymmetry in the partial width one has
\beq
\frac{\Gamma (D^+ \to f)  - \Gamma (D^- \to \bar f)}{\Gamma (D^+ \to f) + \Gamma (D^- \to \bar f) } 
= - \frac{2{\rm sin}(\Delta \phi_W) {\rm sin}(\Delta \delta^{\rm FSI})|{\cal A}_2/{\cal A}_1|} 
{1+|{\cal A}_2/{\cal A}_1|^2 +2|{\cal A}_2/{\cal A}_1|
{\rm cos}(\Delta \phi_W) {\rm cos}(\Delta \delta^{\rm FSI})|  } 
\eeq
with $\Delta \phi_W = \phi_1^{\rm weak} -\phi_2^{\rm weak}$ \& 
$\Delta \delta ^{\rm FSI} = \delta_1^{\rm FSI} -  \delta_2^{\rm FSI}$. We see clearly how
\cp\ asymmetries arise when there are  differences in both weak \& strong phases.

However, the constraints from \cpt \ invariance are not apparent. Suppose the decay mode $f$
belongs to a family of $n$ final states ${f_n}$ connected to each other via re-scattering.
The consequences of \cpt \ invariance 
(general comments on  \cpt \ invariance are given in 
Ref.\cite{JARLSBOOK,WOLFPAPER,KOLYA92,CPBOOK,BGR,BED8})
become visible, if we rewrite the decay amplitude in the form
\bea
T(D^+ \to f_j) &=& e^{i\delta_{f_j}} \left[ T_{f_j} + 
\sum_{f_j \neq f_k}T_{f_k}iT^{\rm resc}_{f_jf_k}  \right] 
\label{CPTAMP1} 
\\
T(D^- \to \bar f_j) &=& e^{i\delta_{f_j}} \left[ T^*_{\bar f_j} +
\sum_{f_k \neq f_j}T^*_{\bar f_j}iT^{\rm resc}_{f_jf_k}  \right]   \; , 
\label{CPTAMP2}
\eea 
where amplitudes $T^{\rm resc}_{f_jf_k}$ describe FSI connecting $f_j$ and  $f_k$ .
One gets, in addition to the direct term, a contribution to the \cp \ asymmetry of the form: 
\beq
\Delta \gamma (a) =  4 \sum_{f_k \neq f_j} T^{\rm resc}_{f_jf_k} \, 
{\rm Im} T^*_{f_j} T_{f_k} 
\label{CPTAMP3}
\eeq

There are subtle, but very important statements about using these equations:
\begin{itemize}

\item 
Final states ${f_n}$ should also include modes with neutrals. In practice, decays like 
$D+ \to \pi^+\pi^0\pi^0$ are really hard to obtain.
\item
\cpt\  invariance can be satisfied in two roads:

\subitem 
One can find that neither $D^{\pm} \to \pi^{\pm}\pi^+\pi^-/\pi^{\pm}\pi^0\pi^0/\pi^{\pm}K^+K^-/
\pi^{\pm}K^0\bar K^0$ shows evidence for \cp\ asymmetry.

\subitem
At least two of them find \cp \ violation with opposite signs. 
\end{itemize}

So far $D^{\pm} \to \pi^{\pm}\pi^+\pi^-$ shows no evidence about \cp\ asymmetry \cite{lhcb3pi}, but  
the two roads are still open.

\subsection{Scattering in $\pi\pi\leftrightarrow  KK$, $\pi\pi\leftrightarrow  \pi \pi$, 
$KK\leftrightarrow  KK$, $\pi K \leftrightarrow \pi K$}
\label{pipiKK}

Early experimental results  from  $\pi\pi$ scattering presented a significant deviation from the elastic regime of the S-wave
in the  region between 1.0-1.5 GeV \cite{CERN_Munich,PENN}.  The inelasticity parameter decreases, starting at 1.0 GeV
get a minimum at 1.2 GeV and come back again to the unitary circle at around 1.5 GeV, going counterclockwise
in the Argand circle. A similar inspection was performed for the P- and D-waves, but no significant deviation from the elastic 
regime was found  in this energy interval. 
 
The deviation of the inelasticity in the S-wave $\pi\pi \to \pi\pi$ scattering is associated to a corresponding 
increase of the cross section of  $\pi\pi\to KK$  \cite{Cohen}, in the same energy region. Notice that due to G-parity conservation
a pair of pions can only scatter into an even number of pions. In other words, a initial state of two pions can produce either
two pions or two kaons.

The same study performed to  $K \pi$ elastic scattering by the LASS experiment \cite{LASS}  showed that both  S- and P-waves  
have an inelasticity parameter close to unity in the Argand circle, up to 1.4 GeV in the P-wave and 1.5 GeV in the S-wave. 
The D-wave is dominated by the resonance $K_2(1430)$ that decay to  $K \pi$  with a branching fraction of about 50\%\cite{PAT}.
Therefore, re-scattering of the $K\pi$ system is  not relevant to this discussion.

The energy range of the $K^+K^-$ pair is  $2m_K\leq m(KK)\leq m_D-m_{\pi}$, which coincides with the range where the 
inelasticity of the $\pi\pi$ scattering deviates from unity. \cpt \ invariance, therefore, connects the
$D^+ \to \pi^+\pi^+\pi^-$  and the   $D^+ \to \pi^+K^+K^-$ decays, through the S-wave $\pi\pi\leftrightarrow KK$ scattering. 
A comprehensive argument should  include $\pi^0\pi^0$ and $K^0\bar K^0$ as well, but this will not be adressed in this paper.

\subsection{Some intriguing results: charmless three-body $B^{\pm}$ decays}
\label{BDECAYS}

Recent LHCb results on charmless three-body $B^{\pm}$ decays show 
sizable averaged \cp\ asymmetries over the FS  with correlations \cite{LHCb028}: 
\bea
A_{CP}(B^{\pm} \to K^{\pm} \pi^+\pi^-) &=&  
+0.032 \pm 0.008_{\rm stat} \pm 0.004_{\rm syst}
\pm 0.007_{\psi K^{\pm}}  
\label{SUPP1}
\\
A_{CP}(B^{\pm} \to K^{\pm} K^+K^-) &=&   
- 0.043 \pm 0.009_{\rm stat} \pm 0.003_{\rm syst}
\pm 0.007_{\psi K^{\pm}} \; .
\label{SUPP2}
\eea 
It is important to note  that these \cp\ asymmetries come with {\em opposite} signs. 

The \cp\ asymmetry was measured across the Dalitz plot and this is the most
interesting result. `Local' \cp\ asymmetries come  also with opposite signs, but 
are much larger: 
\bea 
A_{CP}(B^{\pm} \to K^{\pm} \pi^+\pi^- )_{\rm `local'} &=& + 0.678 \pm 0.078_{\rm stat} 
\pm 0.032_{\rm syst}
\pm 0.007_{\psi K^{\pm}} 
\label{SUPP3}
\\
A_{CP}(B^{\pm} \to K^{\pm} K^+K^- )_{\rm `local'} &=& 
- 0.226 \pm 0.020_{\rm stat} \pm 0.004_{\rm syst}
\pm 0.007_{\psi K^{\pm}} \; .
\label{SUPP4}
\eea 
`Local' \cp\ asymmetries mean here: 
\begin{itemize}
\item 
positive asymmetry at low $m_{\pi ^+\pi ^-}$ just below $m_{\rho^0}$; 
\item 
negative asymmetry both at low and high $m_{K^+K^-}$ values. 
\end{itemize} 

There is another important aspect; asymmetries are observed in regions of the phase space 
not associated to any particular resonance.

A very similar effect was observed in
even more CKM suppressed three-body FS, namely $B^+ \to \pi^+\pi^-\pi^+$ and  $B^+ \to \pi^+K^-K^+$.
LHCb experiment has measured these {\em averaged} and `local' \cp\ asymmetries \cite{LHCbPAPER13018}: 
\bea
A_{CP}(B^{\pm} \to \pi^{\pm} \pi^+\pi^-) &=& + 0.120 \pm 0.020_{\rm stat} \pm 
0.019_{\rm syst} \pm 0.007_{J/\psi K^{\pm}}
\label{PPP1}
 \\
A_{CP}(B^{\pm} \to \pi^{\pm} K^+K^-) &=& - 0.153 \pm 0.046_{\rm stat} \pm 
0.019_{\rm syst} \pm 0.007_{J/\psi K^{\pm}}
\label{PPP2} 
\eea
\bea
A_{CP}(B^{\pm} \to \pi^{\pm} \pi^+\pi^-)|_{\rm `local'} &=&  
+0.584 \pm 0.082_{\rm stat} \pm 0.027_{\rm syst}
\pm 0.007_{\psi K^{\pm}}  
\label{SUPP7} \\
A_{CP}(B^{\pm} \to \pi^{\pm} K^+K^-)|_{\rm `local'} &=&   
- 0.648 \pm 0.070_{\rm stat} \pm 0.013_{\rm syst}
\pm 0.007_{\psi K^{\pm}} \; .
\label{SUPP8}
\eea 
Again it is very interesting that LHCb data in 
Eqs.(\ref{PPP1},\ref{PPP2},\ref{SUPP7},\ref{SUPP8})  
show \cp\ asymmetries with {\em opposite} signs  -- as `natural' by \cpt\ invariance, no matter what  forces produce them. 
Again, a  \cpt\ symmetry argument has to include neutral bosons as well. 

In summary, the results from charmless three-body $B^{\pm}$ decays are very intriguing. Large 
regional effects, diluted 
when phase space integration is performed, appear in regions not associated to resonances, and with opposite signs
in FS that are related by re-scattering. Do they show impact of ND? We refer to 
\cite{BGR,BED8,IKFPCP13,ROSNER} for additional
discussions on this issue.

\section{Simulations of $D^{\pm} \to \pi^{\pm} K^-K^+$  and  $D^{\pm} \to \pi^{\pm} \pi^-\pi^+$}

\subsection{Correlations between $D^{\pm} \to \pi^{\pm}K^+K^-$ and 
$D^{\pm} \to\pi^{\pm}\pi^+\pi^-$}

In this section we perform simulations of the $D^{\pm} \to \pi^{\pm} K^-K^+$ Dalitz plot to
illustrate the re-scattering  effects discussed above. The simulations are performed in the
framework of the isobar model. It is well known that a sum of Breit-Wigners plus a nonresonat
term is not a correct representation of the S-wave\cite{MEISSNER}, but the goal here is not
to extract quantitative information on the resonant structure of the decay. Rather, we are interested
in the differences between $D^+$ and $D^-$ Dalitz plots that reflect CPV effects with and without
the  constraints of \cpt\ constraint. 

For the decay amplitude we use the  resonant substructure is based on Dalitz plot analysis performed 
by CLEO-c collaboration \cite{CLEO}. For simplicity, we neglect contributions from amplitudes that result 
in decay fractions smaller than 1\%. The resonant amplitudes are written as a product of form factors, 
relativistic Breit-Wigners and spin amplitudes. We use the following amplitudes: 
$A_{\phi \pi}=A(D^+\to \phi \pi^+)$, 
$A_{K*K}=A(D^+\to K^*(892)^0 K^+)$, 
$A_{K^*_0K}=A(D^+\to K_0^*(1430)^0 K^+)$, 
$A_{a_0\pi}=A(D^+\to a_0(1450)^0 \pi^+)$, and
$A_{\kappa K}=A(D^+\to \kappa(800) K^+)$.

The decay amplitudes are written as
\bea
\mathcal{A} &=& \sum c_j A_j
\hskip .5cm 
\\
\bar \mathcal{A} &=& \sum \bar c_j A_j
\hskip .5cm
\label{isobar}
\eea
with $c_j\equiv a_je^{i\delta_j}$, $j = \phi \pi,K*K,K^*_0K, a_0\pi, \kappa K$. The
amplitudes $A_j$ involve only {\em CP}-even, strong phases from the Breit-Wigner functions.
Weak phases are included in the phase of the $c_j$ coefficients. \cp \ conservation imply  
$c_j=\bar c_j$ for all $j$.

\begin{figure}[htpb!]
\centering{\includegraphics[width=8cm]{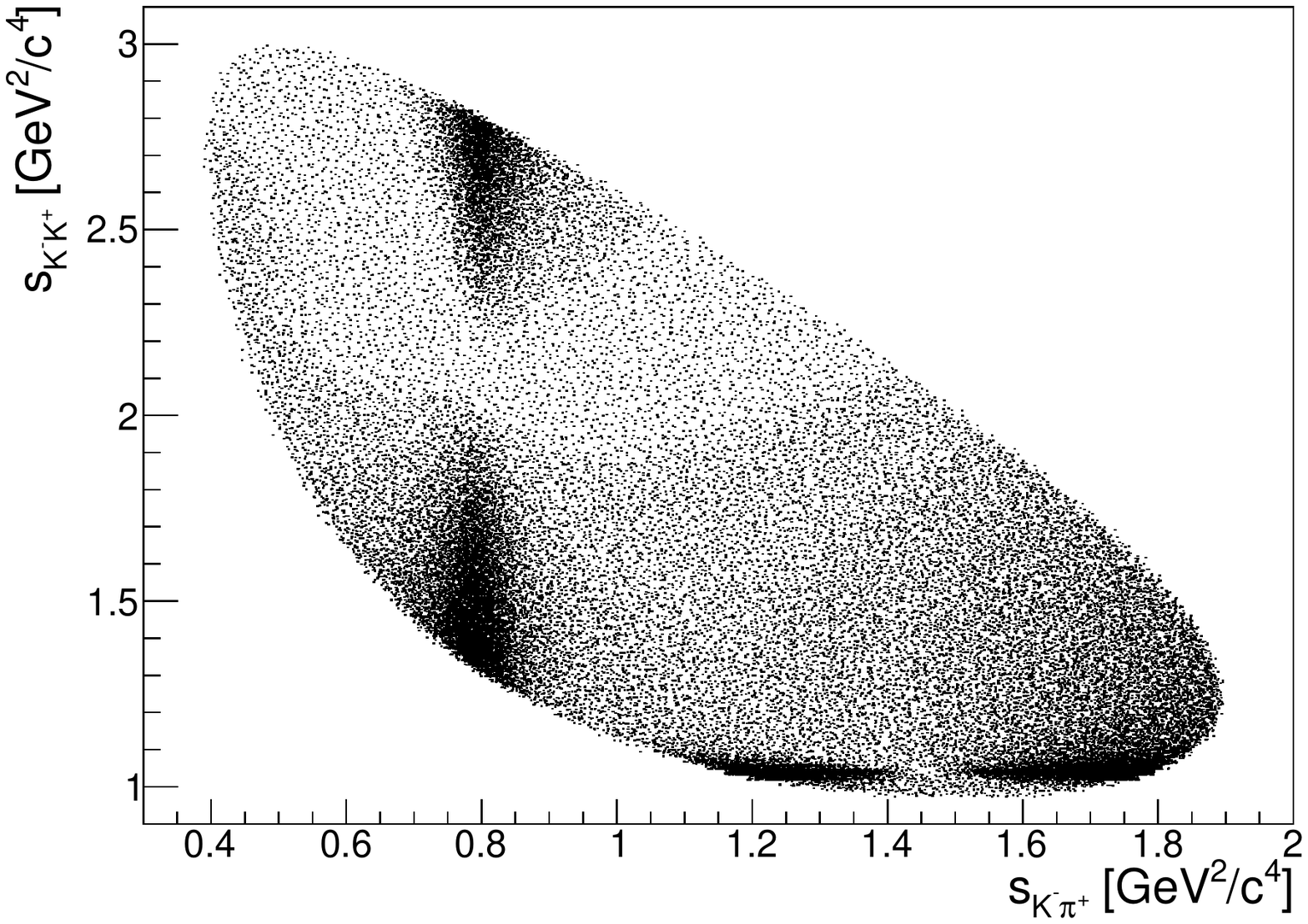}}
\caption{ A simulation of the Dalitz plot of the decay $D\to K^-K^+\pi^+$. The decay model
is taken from CLEO-c (see text for details) and is used as the starting point of our studies.}
\label{dpm}
\end{figure}

The couplings $c_j$   between the $j-th$ resonant mode and the 
initial state are complex for two reasons: 
\begin{itemize}
\item 
Weak forces between quarks may produce phases that are opposite for anti-quarks. 
\item
the decay amplitude is affected by hadronic FSI. Strong phases due to the resonance-bachelor 
re-scattering are included in $\delta_j$ and they are the same for hadrons and anti-hadrons. 
\end{itemize} 
The Dalitz plot of the $D^{\pm} \to \pi^{\pm} K^-K^+$ decay is shown in Fig. \ref{dpm}. The prominent
contributions from the $\phi \pi$ and $K^*(892)^0 K^+$ are clearly visible. The contribution  from the broad 
S-wave $K^-\pi^+$ resonances can be seen at the edges of the $s_{K^-\pi^+}$ axis.

The set of coefficients $c_j$ ($\bar c_j$) defines, thus, the decay
amplitude $\mathcal{A}$ ($\bar \mathcal{A}$). In our simulations we
assume no production asymmetries and identical detection efficiencies, so 
the number of $D^+$ and $D^-$ decays is proportional to the 
integral of the decay amplitudes over the Dalitz plot, 
\begin{eqnarray}
N_{D^+} &\propto& \int \mid\mathcal{A} \mid^2 ds_{KK}ds_{K\pi}\nonumber\\
N_{D^-} &\propto& \int \mid\bar \mathcal{A} \mid^2 ds_{KK}ds_{K\pi}.
\end{eqnarray}

In the case of \cp \ conservation we have exactly the same number of $D^+$ and $D^-$. But
if there is CPV the values of the two integrals will differ, in general. We simulate  the $D^+$ and 
the $D^-$ Dalitz plot separately, seeding CPV in the latter. We always simulate  
$3 \times 10^6$ $D^+ \to K^-K^+\pi^+$ decays. The  number of generated  $D^-$ decays
is defined according to the ratio of the above integrals, which depends on how CPV is seeded.

The averaged CP asymmetry is computed as:
\begin{equation}
A_{CP} = 
\frac{\int \mid\mathcal{A} \mid^2 ds_{KK}ds_{K\pi}-\int \mid\bar \mathcal{A} \mid^2 ds_{KK}ds_{K\pi}}
{\int \mid\mathcal{A} \mid^2 ds_{KK}ds_{K\pi}+\int \mid\bar \mathcal{A} \mid^2 ds_{KK}ds_{K\pi}}
\end{equation}

The $D^+$ and $D^-$ Dalitz plots, simulated as described above,
are compared using the 'Miranda' method \cite{MIRANDA1,MIRANDA2}.
In this method the  $D^{\pm}$ Dalitz plot is divided into bins; a comparison between
the $D^+$ and $D^-$ Dalitz plot is performed directly in a bin-by-bin basis, computing, for each bin, 
the anisotropy variable 
\beq
\mathcal{S}^i_{CP} = \frac{N^+_i-N^-_i}{\sqrt{N^+_i+N^-_i}},
\eeq
with $N^+_i$ and $N^-_i$ being the $i-th$ bin content of the $D^+$ and $D^-$ Dalitz plots, respectively.

\begin{figure}[htpb!]
\includegraphics[width=8cm]{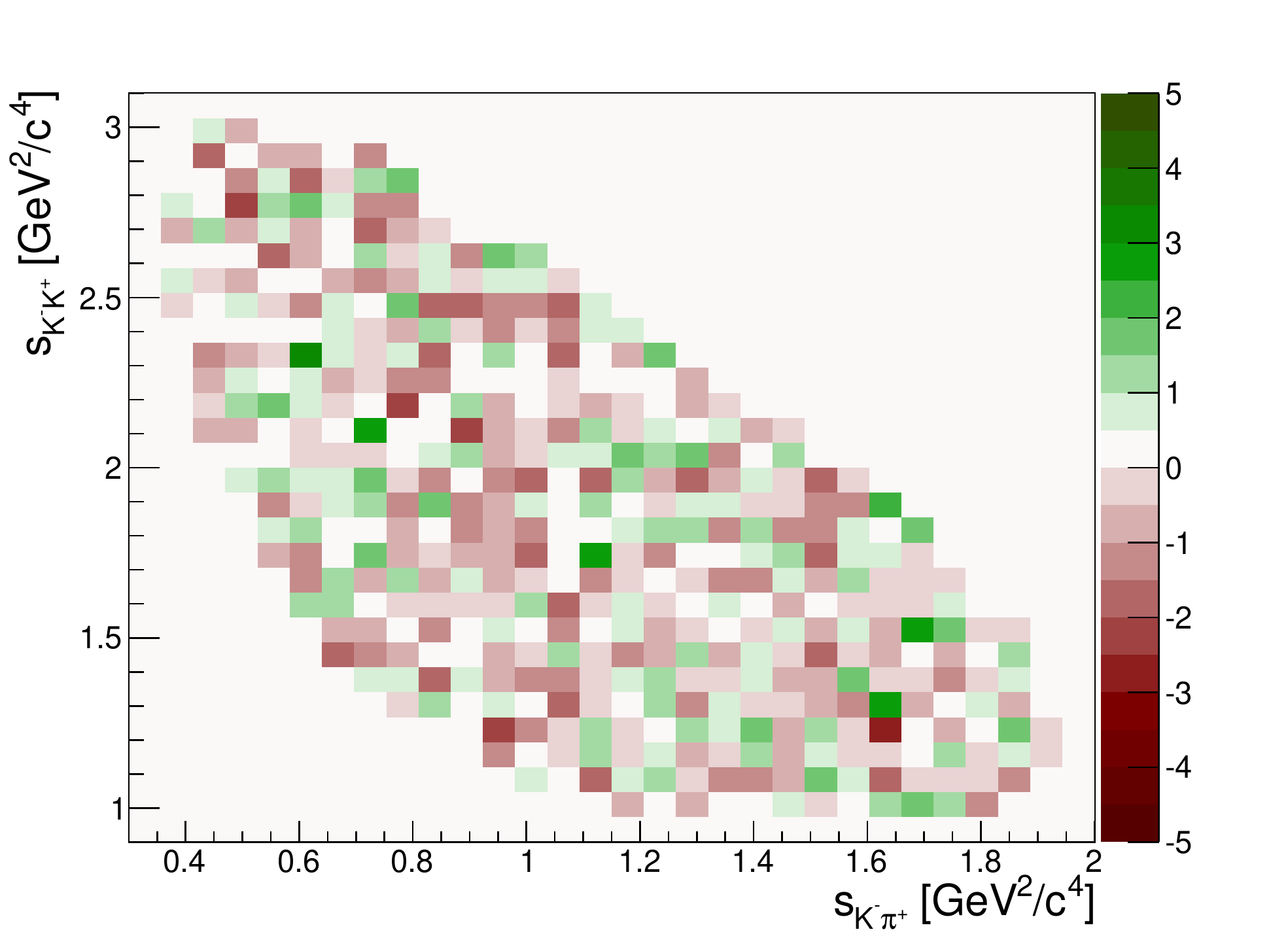}
\includegraphics[width=8cm]{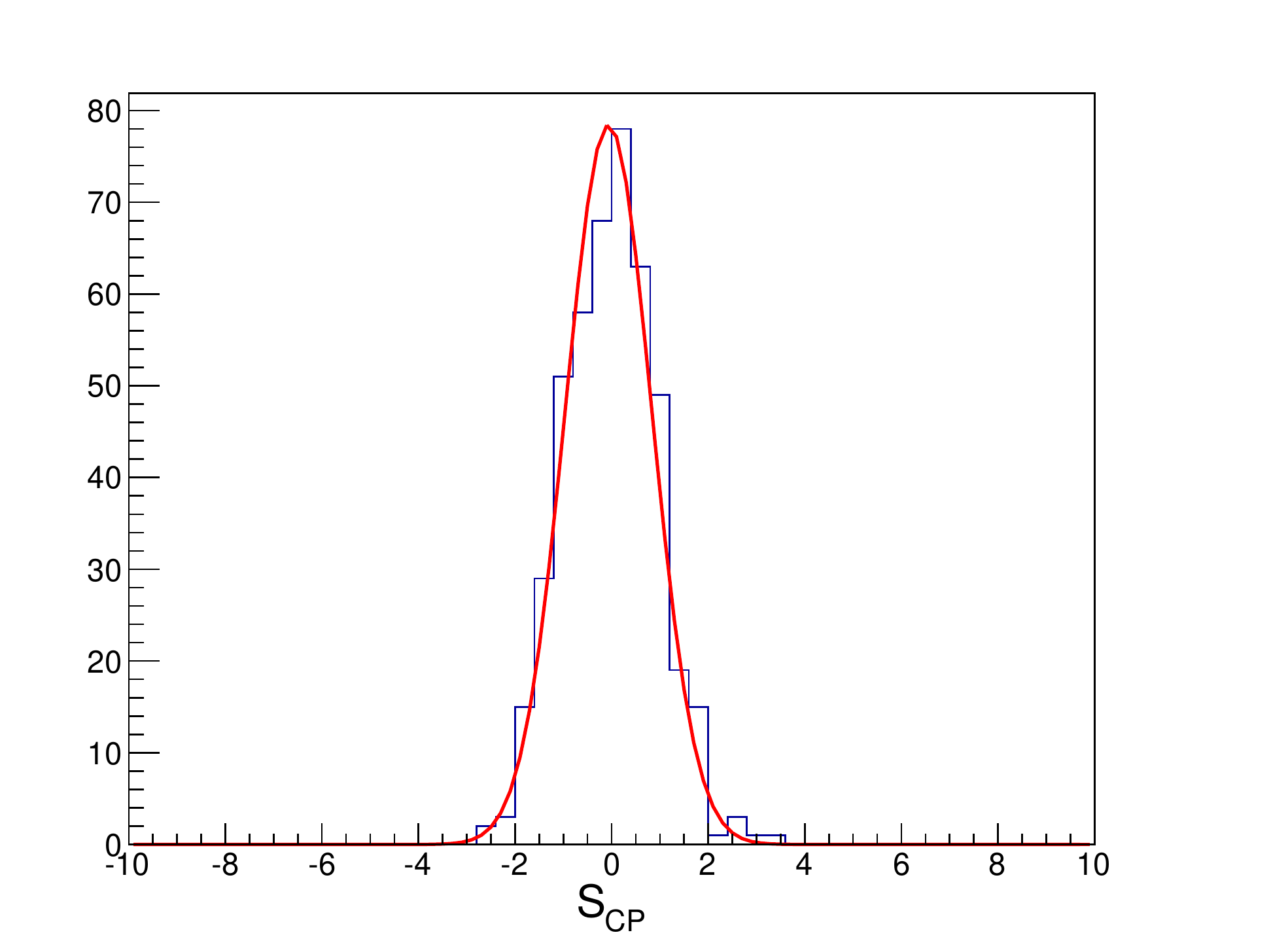}
\caption{ A simulation of the Dalitz plot of the decays $D^+ \to K^-K^+\pi^+$
and $D^- \to K^+K^-\pi^-$. No CPV is seeded: the same set of coefficients $c_j$ is used for the
simulation of the $D^+$ and the $D^-$ samples. Values of $\mathcal{S}_{CP}^i$ are, in this case,
distributed according to a unit Gaussian centered at zero. No
excess of one charge over the other is obeserved in any region of the Dalitz plot, apart from statistical
fluctuations.}
\label{nocpv}
\end{figure}

The value of $\mathcal{S}_{CP}^i$ is a measure of the significance of the excess of one 
charge especie over the other in the $i-th$ bin. Notice that $\mathcal{S}_{CP}^i$ may be positive or negative. 
If  \cp \ is conserved, $N^+_i$ and $N^-_i$ will differ only by statistical fluctuations. The values of
$\mathcal{S}_{CP}^i$,in this case, are distributed according to a unit Gaussian centered at zero.
As an example, we show in Fig. \ref{nocpv} a simulation in which \cp \ is conserved --- the same number 
of $D^+$ and $D^-$ decays are simulated with $c_j = \bar c_j$. The plot on the left 
has the distribution of $\mathcal{S}_{CP}^i$ across the Dalitz plot. No region show any excess of on charge 
over the other, as expected.  The distribution of $\mathcal{S}_{CP}^i$ is shown on the plot on the left, with a unit
Gaussian centered at zero superimposed.

There is a number of models for CPV beyond the SM. In this exercise we assume a simple scenario,
consistent with the SM, in which CPV manifests as a difference in relative phase of the
$K^-\pi^+$ and $K^-K^+$ resonances  in the $D^+$ and the $D^-$ Dalitz plots.  We refer to this as the SM scenario. 
We first simulate CPV  in this  'SM scenario' (SM CPV, for short). Then we simulate the contribution to $D^+ \to K^-K^+\pi^+$ 
from the $D^+ \to \pi^-\pi^+\pi^+$  decay via $\pi^+\pi^- \to K^+ K^-$ re-scattering In this simulation we 'turn off' the SM CPV
and introduce a small CPV effect in $D^+ \to \pi^-\pi^+\pi^+$. Finally the full 
simulation including both effects is performed.

\begin{figure}[htpb!]
\includegraphics[width=8cm]{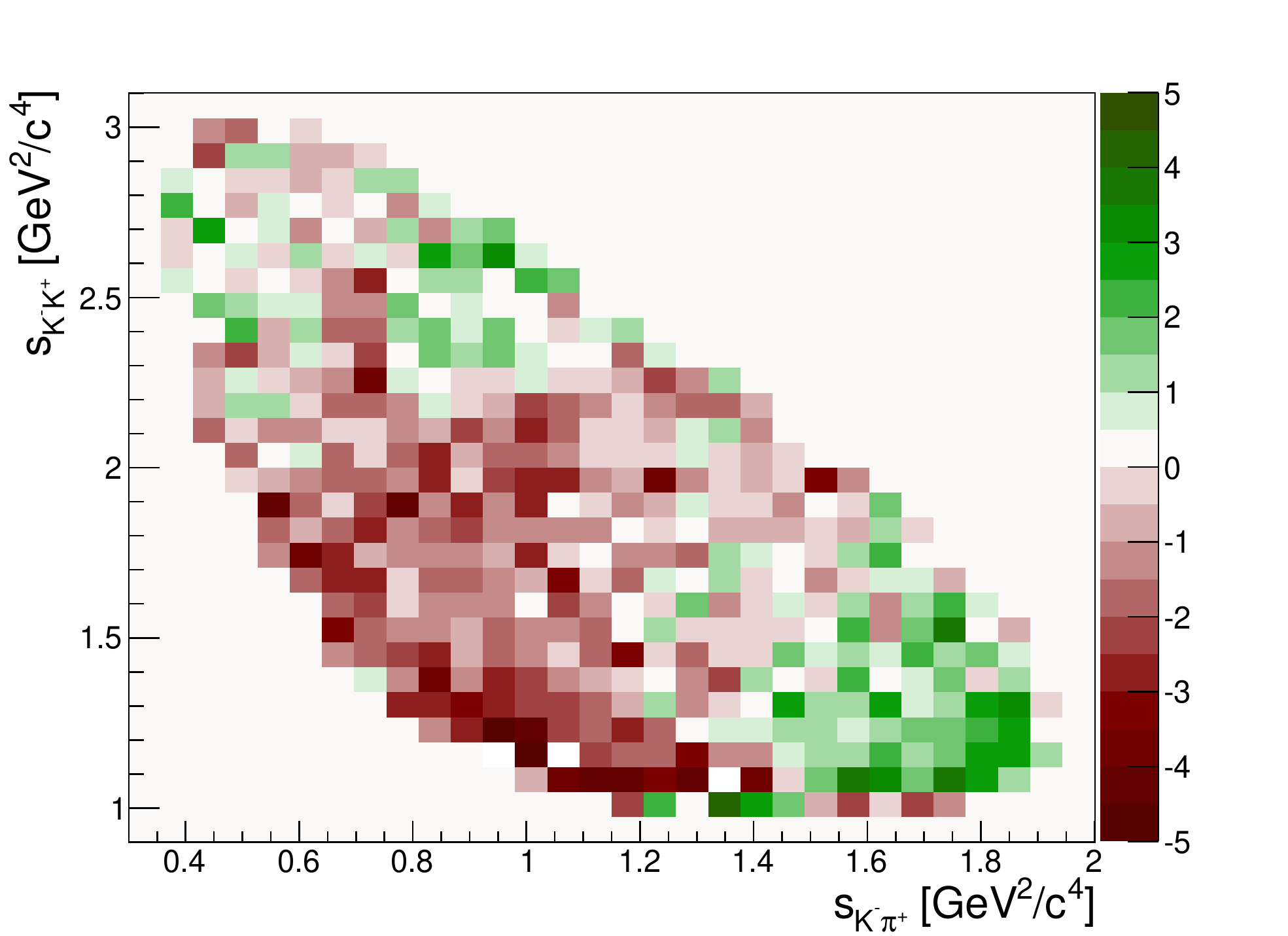}
\includegraphics[width=8cm]{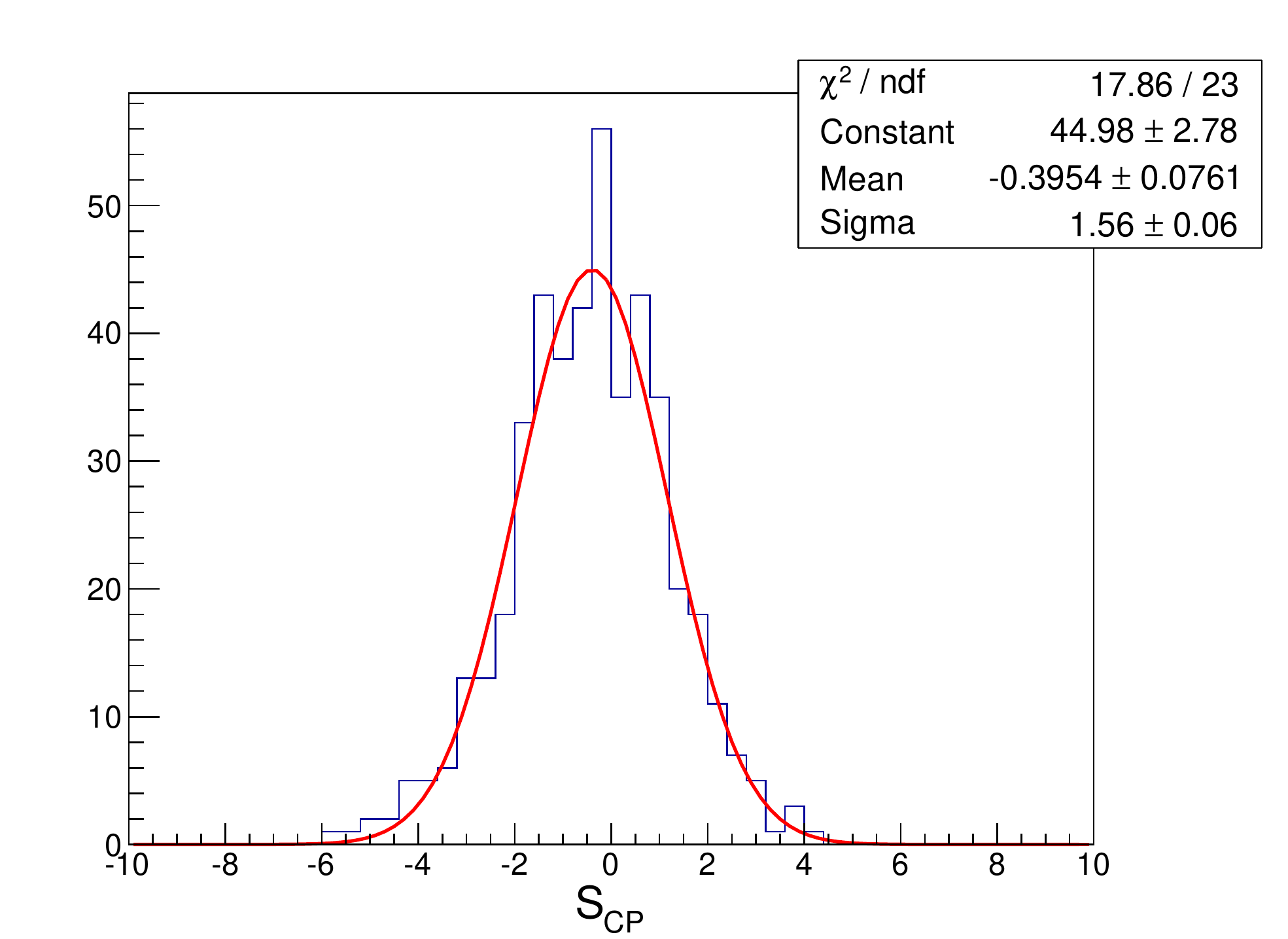}
\caption{ A simulation of {\em CP} violation in the decay $D\to K^-K^+\pi^+$. A 3$^{\circ}$
difference in the $K^*K^+$ and $\phi \pi^+$ relative phase between $D^+$  and $D^-$ is
introduced. The difference in relative phase cause the {\em CP} asymmetry to change sign
across the Dalitz plot, according to the phase variation of the interfering Breit-Wigners functions.}
\label{smcpv}
\end{figure}

Our  SM CPV consists in introducing a 3$^{\circ}$ difference  in the relative phases of the $K^-K^+$ and $K^- \pi^+$ 
resonances when the $D^-$ sample is generated. This 3$^{\circ}$ difference causes a minor excess of $D^-$ over $D^+$ 
resulting in an averaged asymmetry of 0.08\%, beyond the current experimental sensitivity.
The one- and two-dimensional distributions of $\mathcal{S}_{CP}^i$ for the SM CPV simulation
are shown in Fig. \ref{smcpv}. Large local asymmetries are observed, mostly in regions where the 
$K^-K^+$ and $K^- \pi^+$ amplitudes overlap.   
The asymmetry is modulated by the strong phase variation of the resonances, leading
to negative values of $\mathcal{S}_{CP}^i$ in some regions of the Dalitz plot
and positive in another ones. We see how large local effects can result in a very small
averaged asymmetry. The distribution of $S_{CP}$ values is no longer
centered at zero ($\mu=-0.395\pm0.076$)
and is significantly wider than a unit Gaussian ($\sigma=1.56\pm0.06$).

We now illustrate the effect of the \cpt \ constraint. In the $D^+ \to K^-K^+\pi^+$ decay amplitude
we now introduce the contribution from the $D\to \pi^-\pi^+\pi^+$ 
decay through through the $\pi\pi \leftrightarrow KK$ scattering, but keeping $c_j=\bar c_j$. Weak phases
are in general obscured by the strong ones, but here is an instance where the
existence of the strong phase favors the observation of small differences in weak phases. 

The re-scattering term is `inspired' in Eqs.(\ref{CPTAMP1}, \ref{CPTAMP2}). For simplicity, 
the weak amplitude for the $D^+ \to \pi^-\pi^+\pi^+$ decay is represented by a complex constant, 
$T_{D\to 3\pi}$, with an unknown modulus and {\em CP} odd phase. 

The $\pi\pi\to KK$ scattering amplitude is written as 
$T_{\pi\pi\to KK}= A_{\pi\pi\to KK} e^{i\delta_{\pi\pi\to KK}}$. The real functions 
$A_{\pi\pi\to KK}$ and $\delta_{\pi\pi\to KK}$ are taken from \cite{Cohen}. $T_{\pi\pi\to KK}$ is {\em CP} invariant.

The decay amplitudes become
\begin{equation}
\mathcal{A} =  c_{\phi\pi} A_{\phi\pi} + c_{a_0\pi} A_{a_0\pi} + c_{\kappa K} A_{\kappa K} + 
c_{K^*K} A_{K^*K} + c_{K^*_0K} A_{K^*_0K} + T_{D\to 3\pi} T_{\pi\pi\to KK}.,
\label{isobar1}
\end{equation}
\begin{equation}
\overline{\mathcal{A}} =  \bar c_{\phi\pi} A_{\phi\pi} + \bar c_{a_0\pi} A_{a_0\pi} +  
c_{\kappa K} A_{\kappa K} + c_{K^*K} A_{K^*K} + \bar c_{K^*_0K} A_{K^*_0K} +
\bar T_{D\to 3\pi} T_{\pi\pi\to KK}.
\label{isobar2}
\end{equation}

Before performing the full simulation, we investigate the effect of the re-scattering term {\em alone},
which means $c_j=\bar c_j$. In Eqs. (\ref{isobar1}, \ref{isobar2}) we set
$\mid \bar T_{D\to 3\pi}\mid = 0.9 \mid T_{D\to 3\pi}\mid$ and 
$\mathrm{arg}(\bar T_{D\to 3\pi})=\mathrm{arg}(T_{D\to 3\pi}) + 5^{\circ}$. The values of 
$\mid T_{D\to 3\pi}\mid$ and $\mathrm{arg}(T_{D\to 3\pi})$ are unknown. We chose arbitrary
values that yield a small decay fraction of approximately 2\% for the re-scattering 
contribution. This small amount of re-scattering and the small difference introduced between 
$T_{D\to 3\pi}$ and $\bar T_{D\to 3\pi}$ are sufficient to yield a {\em CP} asymmetry of 
approximately 0.7\%, well within the current experimental sensitivity.

\begin{figure}[htpb!]
\includegraphics[width=8cm]{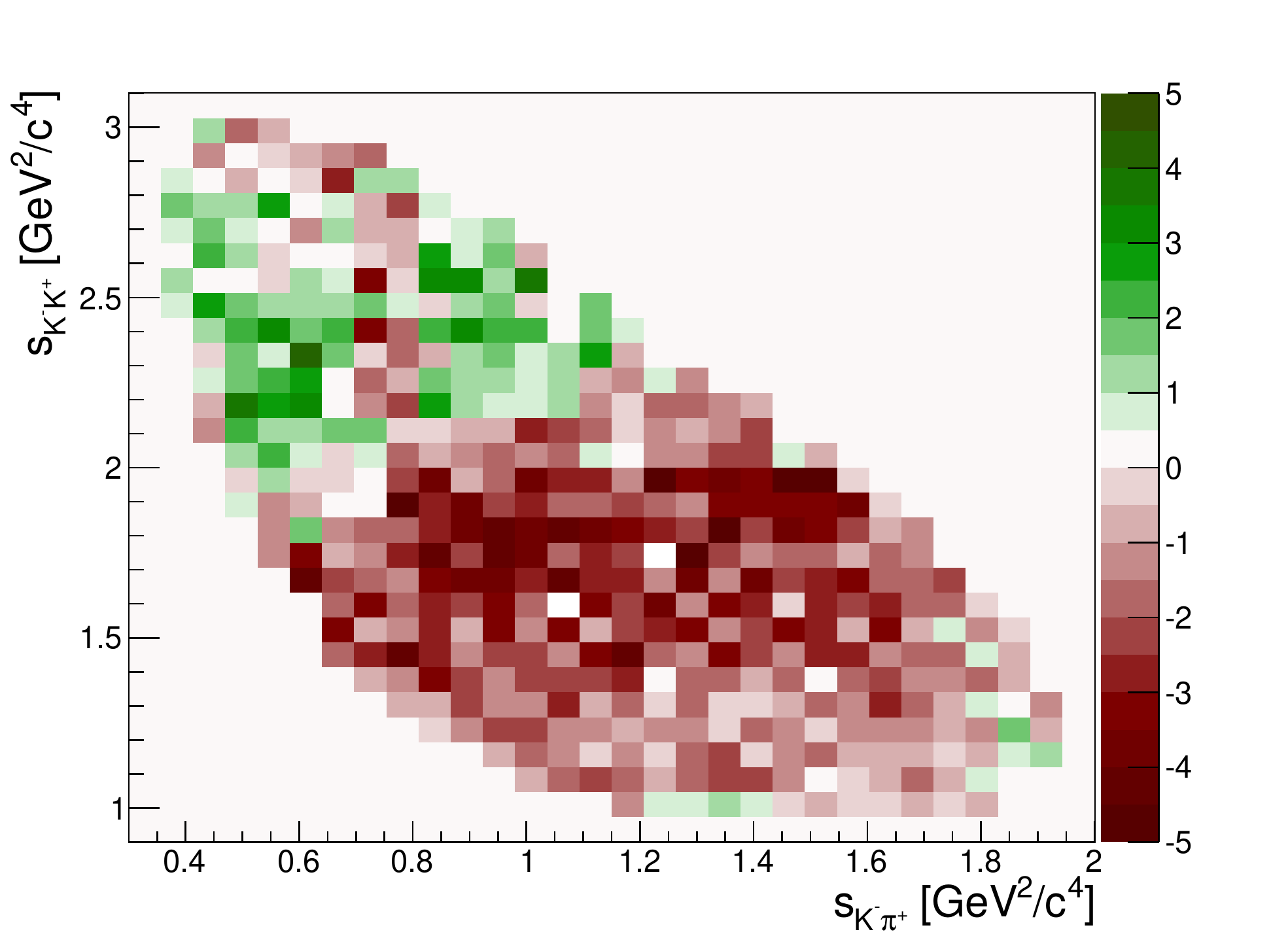}
\includegraphics[width=8cm]{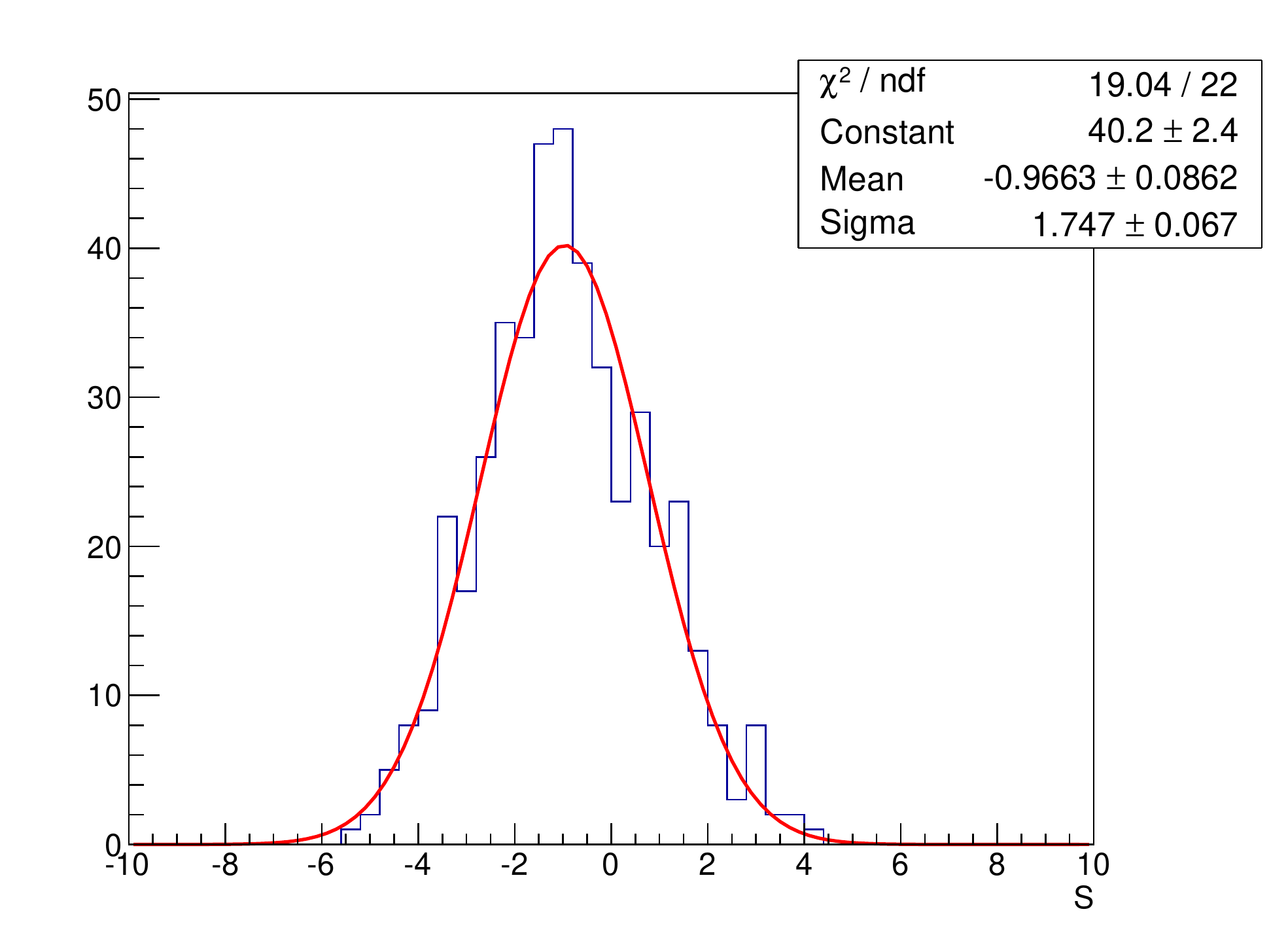}
\caption{ A simulation of the Dalitz plot of the decay $D^+\to K^-K^+\pi^+$. The same set of coefficients $c_j$ are used 
for both $D^+$ and $D^-$. A re-scattering term in the decay amplitude is introduced (see text for details), 
being different for  $D^+$ and $D^-$. The distribution in the left panel is fitted to a Gaussian with free mean and width.}
\label{rescat}
\end{figure}

The one- and two-dimensional distributions of $\mathcal{S}_{CP}^i$ for this simulation
are shown in Fig. \ref{rescat}. The effect of the global asymmetry is
a displacement of the mean of the $\mathcal{S}_{CP}^i$ distribution (right plot).
The widht of the Gaussian, $\sigma=1.747\pm0.067$,  deviates significantly from unity.
The Dalitz plot  exhibits a clear excess of $D^-$ over $D^+$ events towards lower 
values of $m_{K^+K^-}^2$, as expected since $\mid T_{\pi\pi\to KK}\mid$ has a maximum
near 1.2 GeV/$c^2$.

\begin{figure}[htpb!]
\includegraphics[width=9cm]{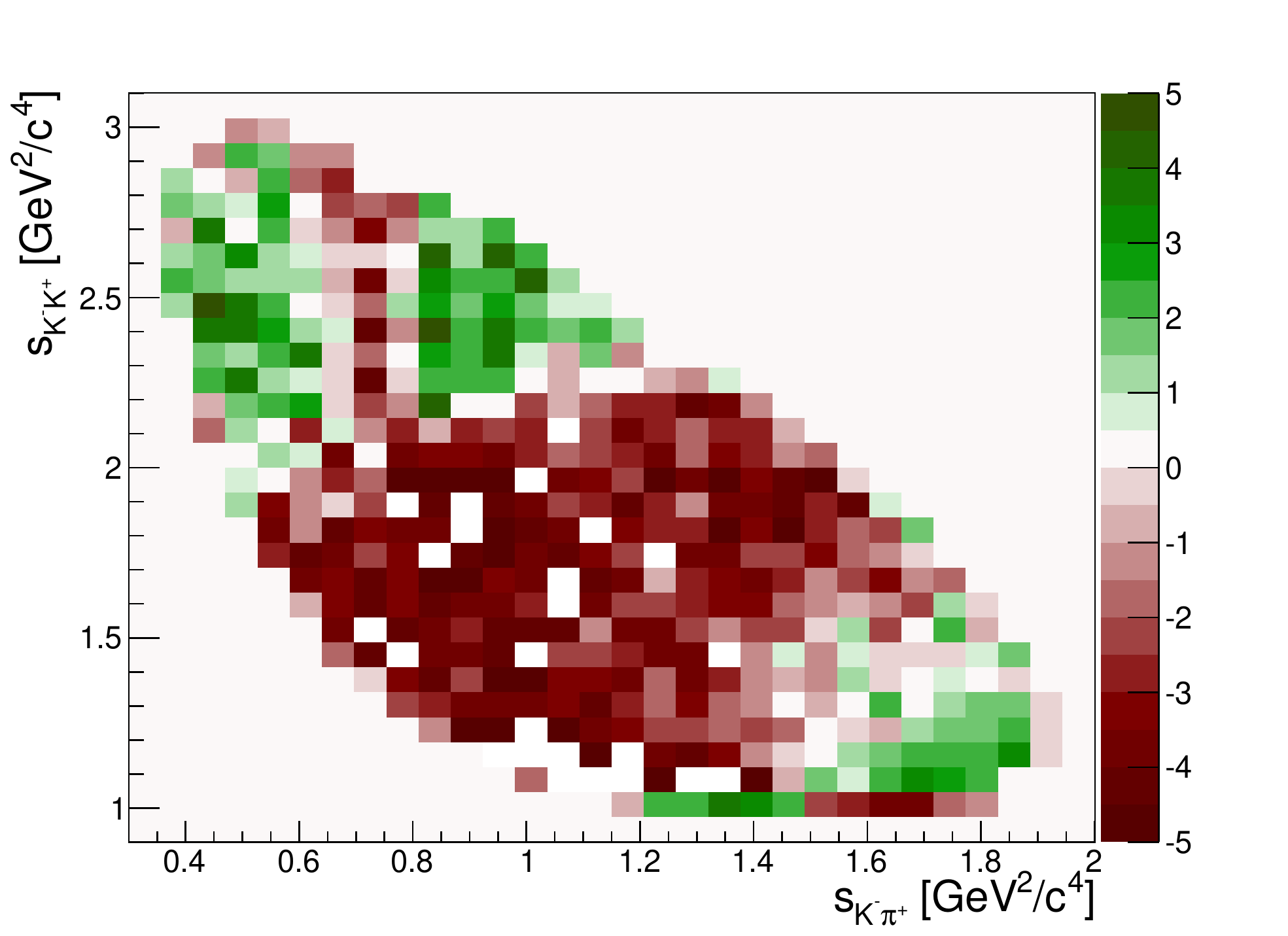}
\includegraphics[width=8cm]{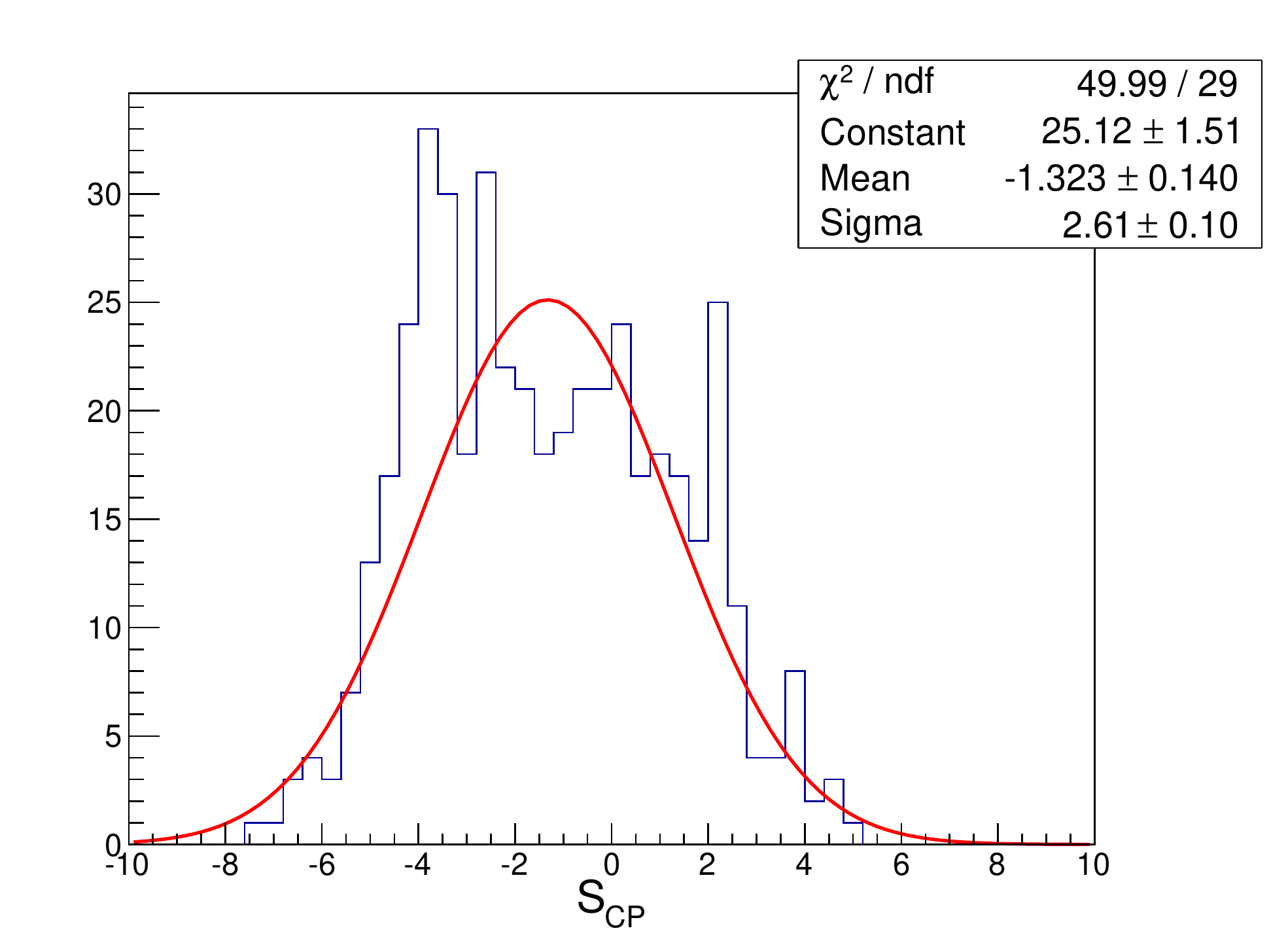}
\caption{ A simulation of {\em CP} violation in the decay $D^+\to K^-K^+\pi^+$. 
A 3$^{\circ}$ difference in the $K^*K^+$ and $\phi \pi^+$ relative phase between $D^+$  and $D^-$ is
introduced, in addition to the difference in the re-scattering term.}
\label{full}
\end{figure}

We are now ready for the full simulation. The $D^-$ sample is generated with the set of
$\bar c_j$ coefficients used in the SM CPV example, whereas the re-scattering term is as
described above. The $\mathcal{S}_{CP}^i$ distributions  are shown in Fig. \ref{full}.

We do not know how large the strong  re-scattering term should be, or what value the weak phase 
of $T_{D\to 3\pi}$ should take.  The effect of the re-scattering in the {\em CP} asymmetry 
depends, naturally, on the assumed difference between $T_{D\to 3\pi}$ and $\overline{T}_{D\to 3\pi}$.
One should keep in mind that decays with neutrals must be considered in a comprehensive treatment.
But with  this simple simulation we show how the addition of a  re-scattering contribution 
may change not only the pattern of the SM-CPV asymmetry of Fig. \ref{smcpv}, but also give rise to a  
global {\em CP} asymmetry. Different combinations of $\mid T_{D\to 3\pi}\mid $ and
$\mathrm{arg}(T_{D\to 3\pi})$ yielding decay fractions up to a few per cent were tested, always with
similar results.  With this investigation we want to call attention to the importance of exploring
the constraints of \cpt \ symmetry,  showing how re-scattering contribution may increase both local and 
phase space integrated effects.

\subsection{ND in $D^{\pm} \to \pi^{\pm}K^+K^-$ with 
$D^{\pm} \to\pi^{\pm}\pi^+\pi^-$}

We discuss now one last example: how one can use the Dalitz plot to access the impact of ND.
There is a number of extensions of the SM.  We explore a scenario in which ND manifests 
as an enhancement of {em CP} violation effects associated with the broad scalars 
(like charged Higgs exchanges). These resonances populates the whole Dalitz plot, interfering 
with all other components. The resulting  asymmetries would be spread all over the phase space.

As discussed before, for the sake of simulations the use of Breit-Wigners parameterization in 
the context of the isobar model is good enough to highlight the impact of CPV. 
Better tools -- like refined dispersion relations \cite{PENN} 
based on the data of low enegy strong scattering -- have to be
developed when it comes to analyse the large data sets from LHCb. 

As in our previous simulations, we use the decay amplitude from CLEO-c \cite{CLEO}, 
for the $D^{\pm} \to \pi^{\pm}K^+K^-$. For the  $D^{\pm} \to \pi^{\pm}\pi^+\pi^-$
we use the results from E791 \cite{e791}.
 {\em CP} violation is seeded as a 1\% difference in the strenght of
the coupling of the  $D^+$ and $D^-$ mesons to the light scalars $\kappa$ and $\sigma$,
plus a 1$^{\circ}$ phase difference.  

\begin{figure}[htpb!]
\includegraphics[width=8 cm]{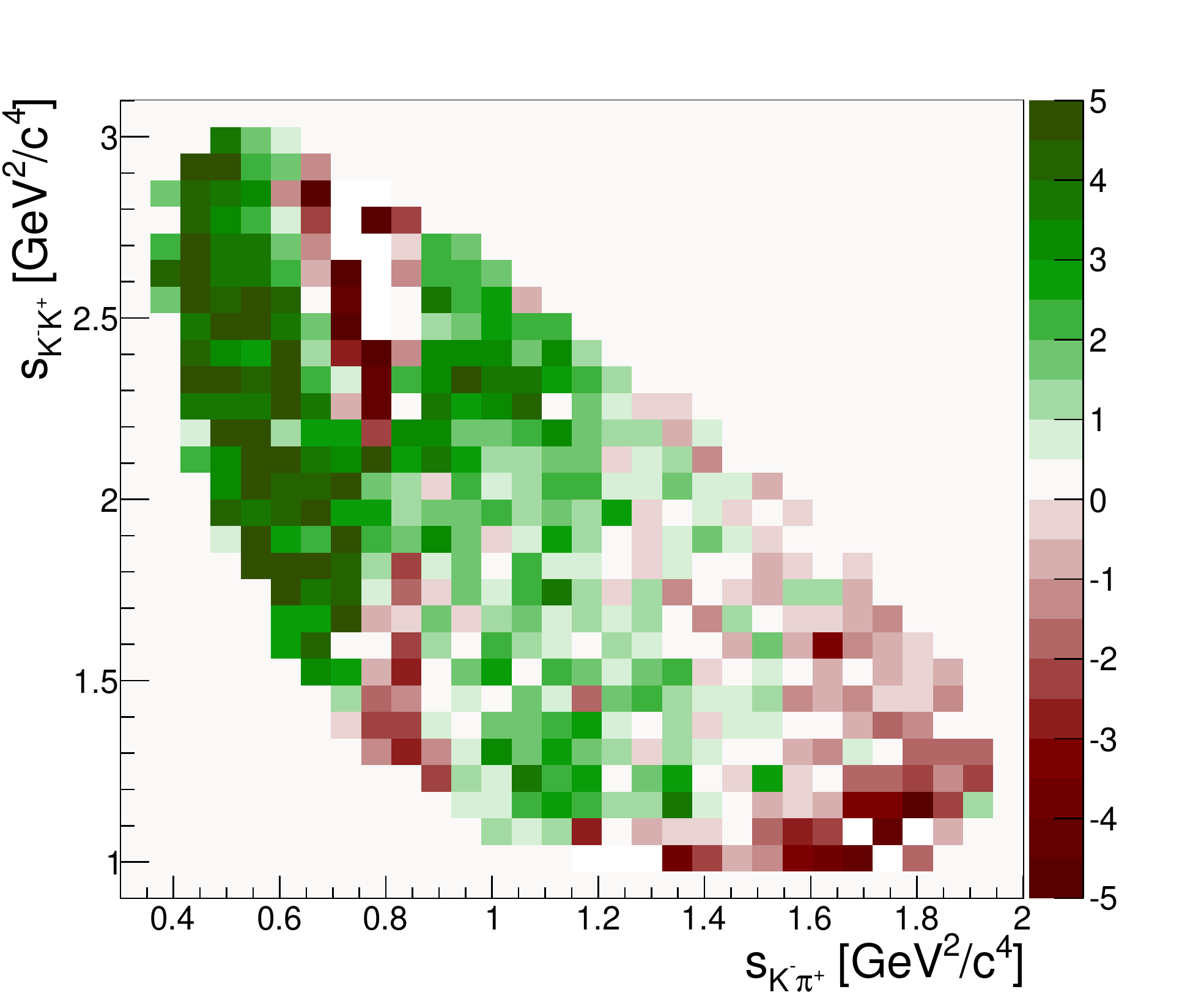}
\includegraphics[width=8 cm]{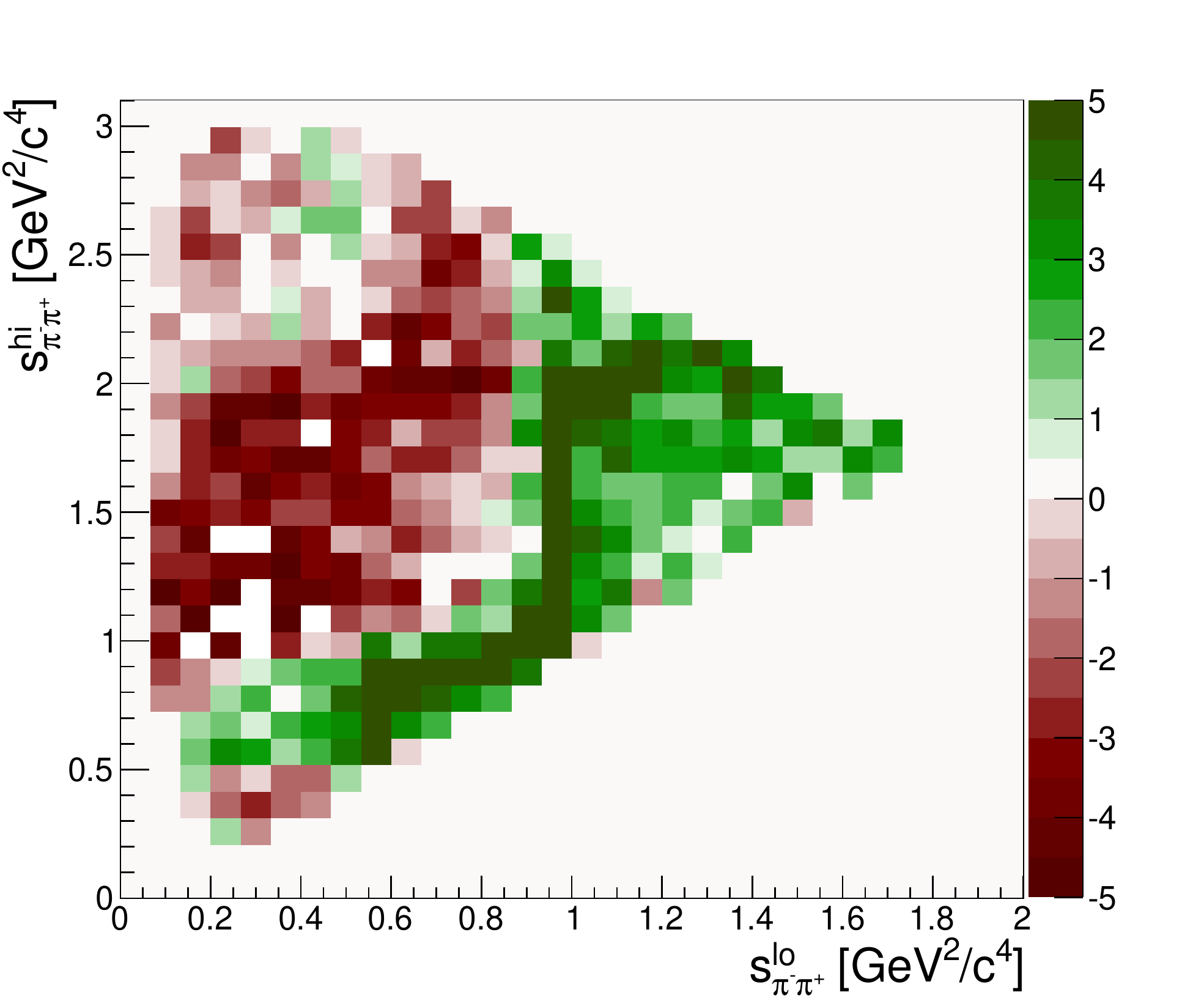}
\caption{Simulation the decays  $D^+\to K^-K^+\pi^+$ 
(left) and $D^+\to \pi^-\pi^+\pi^+$ {right}.  {\em CP} violation is seeded inspired in a ND scenario
in which there is an asymmetry between the coupling between the $D$ meson and the ligth scalars . }
\label{nd}
\end{figure}

The distributions of the values from the Miranda procedure across the Dalitz plot with CPV seeded as 
described above are shown in Fig. \ref{nd}. The broadness of the scalars cause the {\em CP} violation 
effects to be spread over a large portions of the Dalitz plots, being more intense as one approaches the
resonance nominal mass. The asymmetry pattern in this example is significantly different from that of
the SM CPV of Fig. \ref{smcpv}.

\section{Discussion}
\label{CON}

In $B$ transitions one has to find {\em non}-leading source of CP violation. 
We had emphasized the need to go beyond the phase space integrated CP asymmetries
and probe regional effects on Dalitz plots of three-body $B$ decays 
\cite{MIRANDA1,MIRANDA2,BGR,BED8,IKFPCP13}. It is crucial to understand the impact of 
$\pi \pi \leftrightarrow \pi \pi$, $K \bar K \leftrightarrow K \bar K$, $K\pi \leftrightarrow K \pi$ 
and more.   

The landscape is very different for charm decays, where no CP violation has been found yet. 
So far theoretical and experimental efforts have  focused mostly on two-body FS of charm mesons.
This is no surprise since two-body decays are much simpler to treat than three-body ones.
However, in order to understand the possible impact of ND  
in an eventual observations of \cp \ violation 
in charm decays, one definetely needs to go beyond the ratio of integrated rates and study the
pattern of regional CPV. {\em This is the main message of this paper.} 
One has to do it in steps to understand the information 
that the data will give us. 

The SM produces only small CP asymmetries in SCS decays and very close to zero in DCS one.
In this respect, the mere observation of CPV in DCS decays would be a strong indication of ND.
DCS rates, however, are very small and very large data sets would be required. 

Singly Cabibbo suppressed decays are much more promising. Very large data sets already exist.
In this paper we have produced simulations of three-body singly Cabibbo suppressed 
$D^{\pm}$ decays. We focused on the $D^{\pm} \to \pi^{\pm}\pi^+\pi^-/\pi^{\pm}K^+K^-$ and
explored the consequences of the \cpt \ invariance. It is crucial to understand the impact of 
$\pi \pi \leftrightarrow \pi \pi$, $\pi\pi \leftrightarrow K \bar K$, $K\pi \leftrightarrow K \pi$ etc. 

This is obviously very challenging. CPV in decays of heavy flavor involves an interplay between the
degrees of freedom at the quark level and long distance effects of low energy hadron physics. One needs to think beyond the simple valence quark diagrams. The $U$-spin symmetry was invented by 
Lipkin \cite{LIPKIN2}. Later it was applied to $B$ decays many times, as one can see in these 
references \cite{GRG}; in \cite{LIPK} it was suggested that one might to deal with U-spin 
violation of the order of 10 - 20 \%.  

As discussed in Ref.\cite{BIANCO} data tell 
us much larger violations in {\em exclusive} decays $D^0 \to K^+K^-$ vs. $D^0 \to \pi^+\pi^-$ and 
$D^0 \to K^+K^-\pi^+\pi^-$ vs. $D^0 \to \pi^+\pi^-\pi^+\pi^-$ -- however much less in the sum of $D$ decays. 

The simulations we performed illustrates the impact of the correlations due to \cpt \ invariance, which
establishes useful connections between different FS related to each other via strong re-scattering.
FSI interactions are indeed a crucial ingredient for any accurate Dalitz plot analysis with the contemporary data sets. Much more theoretical work is necessary in order to produce better decay models.

\vspace{0.5cm}

{\bf Acknowledgments:} This work was supported by the NSF under the grant numbers PHY-1215979 and by CNPq.

\vspace{4mm}


\end{document}